\documentclass[3p,draft,number]{elsarticle}

\usepackage{mathpazo,times,mathrsfs,amsmath,amsfonts,amsthm,paralist}
\usepackage[ruled,noend,nofillcomment,linesnumbered]{algorithm2e}
\SetAlTitleFnt{}\SetAlgoInsideSkip{bigskip}\SetArgSty{}\SetCommentSty{textit}\AlCapSkip 1.7ex
\usepackage[all,dvips,color,arrow,curve]{xy}\xyoption{arc}

\newtheorem{theorem}{Theorem}
\newtheorem{corollary}[theorem]{Corollary}
\newtheorem{lemma}[theorem]{Lemma}
\theoremstyle{definition}
\newtheorem{definition}[theorem]{Definition}
\newtheorem{fact}[theorem]{Fact}

\renewcommand{\qed}{\raisebox{0.8ex}{\framebox{}}}
\renewenvironment{proof}{\noindent{\bf Proof.}\ }{\hfill\qed\par\bigskip}
\newenvironment{proofof}[1]{\noindent{\bf Proof of #1.}\ }{\hfill\qed\par\bigskip}

\newcommand{\vl}{\ensuremath{v\ell}} 
\newcommand{\T}{\ensuremath{\mathscr{T}}} 
\newcommand{\C}{\ensuremath{\mathscr{C}}} %
\newcommand{\Hh}{\ensuremath{\mathscr{H}}} %
\newcommand{\Ee}{\ensuremath{\mathscr{E}}} %
\newcommand{\Ll}{\ensuremath{\mathscr{L}}}
\newcommand{\Sum}{\displaystyle\sum}

\newcounter{claim}
\newenvironment{claim}[1][]{\refstepcounter{claim}\vspace{1ex}\noindent{(\it\arabic{claim}) {#1}{}}\it}{\vspace{1ex}}
\newenvironment{proofclaim}[1][]{\noindent {}{#1}{}}{This proves~(\arabic{claim}).\vspace{1ex}}

\begin{document}

\title{The vertex leafage of chordal graphs\tnoteref{thk}}
\author[wlu]{Steven Chaplick}\ead{chaplick@cs.toronto.edu}
\author[warwick]{Juraj Stacho}\ead{j.stacho@warwick.ac.uk}
\address[wlu]{Department of Physics and Computer Science, Wilfrid Laurier
University, 75 University Ave. West, Waterloo, Ontario N2L 3C5, Canada}
\address[warwick]{DIMAP and Mathematics Institute, University of Warwick,
Coventry CV4 7AL, United Kingdom}

\tnotetext[thk]{The main work on this project was done during the second
author's visit at the University of Toronto in 2009 and later during the first
author's visit at the Caesarea Rothschild Institute of the University of Haifa
in 2011. Both visits were supported by Prof. Derek Corneil of the University of
Toronto via his NSERC grant. The second author also gratefully acknowledges
support from EPSRC, award EP/I01795X/1.}


\begin{abstract}
Every chordal graph $G$ can be represented as the intersection graph of a
collection of subtrees of a host tree, a so-called {\em tree model} of $G$. The
leafage $\ell(G)$ of a connected chordal graph $G$ is the minimum number of
leaves of the host tree of a tree model of $G$.  The vertex leafage $\vl(G)$
is the smallest number $k$ such that there exists a tree model of $G$ in which
every subtree has at most $k$ leaves.  The leafage is a polynomially computable
parameter by the result of \cite{esa}.  In this contribution, we study the
vertex leafage.

We prove for every fixed $k\geq 3$ that deciding whether the vertex leafage of a
given chordal graph is at most $k$ is NP-complete by proving a stronger
result, namely that the problem is NP-complete on split
graphs with vertex leafage of at most $k+1$. On the other hand, for chordal
graphs of leafage at most $\ell$, we show that the vertex leafage can be
calculated in time $n^{O(\ell)}$.  Finally, we prove that there exists a tree
model that realizes both the leafage and the vertex leafage of $G$. Notably, for
every path graph $G$, there exists a path model with $\ell(G)$ leaves in the
host tree and it can be computed in $O(n^3)$ time.
\end{abstract}

\maketitle

\section{Introduction}
In the following text, a graph is always finite, simple, undirected, and
loopless. We write that a graph $G=(V,E)$ has vertex set $V(G)$ and edge set
$E(G)$.  We write $uv$ for the edge $(u,v)\in E(G)$. We write $N_G(v)$ to denote
the neighbourhood of $v$ in $G$, and write $N_G[v]=N_G(v)\cup\{v\}$. The degree
of $v$ in $G$ is denoted by $\deg_{G}(v)=|N_G(v)|$. Where appropriate, we drop
the index $G$, and write $N(v)$, $N[v]$, and $\deg(v)$, respectively. We write
$G[X]$ to denote the subgraph of $G$ induced by $X\subseteq V(G)$, and write
$G-X$ for the graph $G[V(G)\setminus X]$. We write $G-v$ for $G-\{v\}$. We say
that $X$ is a {\em clique} of $G$ if $G[X]$ is a complete graph, and $X$ is an
{\em independent set} of $G$ if $G[X]$ has no edges.

A {\em tree model} of a graph $G=(V,E)$ is a pair $\T=(T,\{T_u\}_{u\in V})$
where $T$ is a tree, called a {\em host tree}, each $T_u$ is a {\em subtree} of
$T$, and a pair $uv$ is in $E$ if and only if $V(T_u)\cap V(T_v)\neq\emptyset$.
In other words, $\T$ consists of a host tree and a collection of its subtrees
whose vertex intersection graph is $G$. 

A graph is {\em chordal} if it does not contain an induced cycle of length four
or more.  It is well-known \cite{buneman,gavril,walter} that a graph is chordal
if and only if it has a tree model.

For a tree $T$, let $\Ll(T)$ denote the set of its {\em leaves}, i.e., vertices of
degree one.  If $T$ consists of a single node, we define $\Ll(T)=\emptyset$.
In other words, we consider such a tree to have no leaves.

The {\em leafage} of a chordal graph $G$, denoted by $\ell(G)$, is defined as
the smallest integer $\ell$ such that there exists a tree model of $G$ whose
host tree has $\ell$ leaves (see \cite{leafage}). It is easy to see that
$\ell(G)=0$ if and only if $G$ is a complete graph, and  otherwise $\ell(G) \geq
2$.  Moreover the case $\ell(G)\leq 2$ corresponds precisely to {\em interval
graphs} (intersection graphs of intervals of the real line)
\cite{fulkersongross}. In this sense, the leafage of a chordal graph $G$
measures how close $G$ is to being an interval graph. 

In this paper, we study a similar parameter. 

\begin{definition}
For a chordal graph $G=(V,E)$, the {\em vertex leafage} of $G$, denoted by
$\vl(G)$, is the smallest integer $k$ such that there exists a tree model
$\big(T,\{T_u\}_{u\in V}\big)$ of $G$ where $|\Ll(T_u)|\leq k$ for all $u\in V$. 
\end{definition}

\noindent In other words, the vertex leafage of $G$ seeks a tree model of $G$
where each of the subtrees (corresponding to the vertices of $G$) has at most
$k$ leaves and the value of $k$ is smallest possible.

As in the case of leafage, the vertex leafage is a natural parameter related to
some subclasses of chordal graphs previously studied in the literature.  We note
that $\vl(G)\geq 2$ unless $G$ is a complete graph (in which case $\vl(G)=0$),
and the case $\vl(G)\leq 2$ corresponds precisely to the so-called {\em path
graphs} (intersection graphs of paths in trees) \cite{gavril-path} (see also
\cite{chaplick,leveque,monmawei,schaeffer-path}). Thus, the vertex leafage of a chordal graph
$G$ can be seen as a way to measure how close $G$ is to being a path graph.
In \cite{linear-leafage}, it is further observed that in $O(kn)$ time one can
find: an optimal colouring, a maximum independent set, a maximum clique, and an
optimal clique cover of an $n$-vertex chordal graph $G$ with vertex leafage~$k$
if a representation of $G$ (a tree model realizing vertex leafage) is given. 

In \cite{gavril-path} it is shown that path graphs can be recognized in
polynomial time. Currently, the best known recognition algorithms for path
graphs run in $O(nm)$ time \cite{chaplick,schaeffer-path}, where $n=|V(G)|$ and
$m=|E(G)|$. In other words, for a graph $G$, testing whether $\vl(G) \le 2$ can
be performed in $O(nm)$ time. 

Some other restrictions/variations on the standard tree model have also been
studied. One such family of these variations is captured by the $[h,s,t]$ graphs
(introduced in \cite{jamison-hst}) defined as follows: $G=(V,E)$ is an $[h,s,t]$
graph if there is a tree model $\big(T,\{T_u\}_{u\in V}\big)$ of $G$ such that
the maximum degree of $T$ is at most $h$, the maximum degree of each of
$\{T_u\}_{u\in V}$ is $s$, and $uv$ is an edge of $G$ if and only if $T_u$ and
$T_v$ have at least $t$ vertices in common. For more information on these graphs
see \cite{golumbic-hst2,golumbic-hst1}.

\medskip
We summarize the results of our paper in the following theorems.

\begin{theorem}\label{thm:1}
For every $k\geq 3$, it is NP-complete to decide, for a split graph $G$ whose
vertex leafage is at most $k+1$, if the vertex leafage of $G$ is at~most~$k$.
\end{theorem}

\begin{theorem}\label{thm:2}
For every $\ell\geq 2$, there exists an $n^{O(\ell)}$ time algorithm that, given
an $n$-vertex chordal graph $G$ with $\ell(G)\leq \ell$, computes the vertex leafage of $G$
and construct a tree model of $G$ that realizes the vertex leafage of $G$.
\end{theorem}

\begin{theorem}\label{thm:3}
There exists an $O(n^3)$ time algorithm that, given an $n$-vertex chordal graph $G=(V,E)$ and
a tree model $(T,\{T_u\}_{u\in V})$ of $G$, computes a tree model
$(T^*,\{T^*_u\}_{u\in V})$ of $G$ such that\smallskip

\begin{compactenum}[(i)]
\item $|\Ll(T^*_u)|\leq |\Ll(T_u)|$ for all $u\in V$,
\item $|\Ll(T^*)|=\ell(G)$.
\end{compactenum}
\end{theorem}

\begin{corollary}\label{cor:4}
For every chordal graph $G=(V,E)$, there exists a tree model
$(T^*,\{T^*_u\}_{u\in V})$  such that\smallskip

\begin{compactenum}[(i)]
\item $|\Ll(T^*_u)|\leq \vl(G)$ for all $u\in V$.
\item $|\Ll(T^*)|=\ell(G)$,
\end{compactenum}\smallskip
In other words, such a tree model
is optimal with respect to the leafage and also the vertex leafage of $G$.
\end{corollary}

This paper is structured as follows. First, in \S \ref{sec:clique-tree}, we
discuss some technical details related to tree models. After that, in \S
\ref{sec:vl_is_npc}, we prove for every fixed $k\geq 3$ that deciding whether
the vertex leafage of a chordal graph is at most $k$ is NP-complete (i.e., we
prove Theorem~\ref{thm:1}). In light of theorem \ref{thm:1}, in \S
\ref{sec:fixed_l}, we discuss calculating vertex leafage subject to bounded
leafage. More specifically, for bounded leafage $\ell$, we show how to compute
the vertex leafage in time $n^{O(\ell)}$ (i.e., we prove Theorem~\ref{thm:2}).
Finally, in \S \ref{sec:fixed_vl}, we show that the vertex leafage and leafage
of any chordal graph $G$ can be realized simultaneously in a tree model of $G$
(i.e., we prove Theorem \ref{thm:3} and Corollary \ref{cor:4}). We close the
paper in \S \ref{sec:conclusion} with a summary and a discussion of possible
extensions of this work.

\section{Minimal Tree Models and Clique Trees}
\label{sec:clique-tree}

We need to discuss a particular type of tree models of chordal graphs.  Most of
this section is rather technical and a reader experienced with tree models can
easily skip this part. However, we include it for completeness as some of the
subtle transformations involved may not be clear to every reader.

Let $G=(V,E)$ be a chordal graph.  We say that two tree models
$\T=(T,\{T_u\}_{u\in V})$ and $\T'=(T',\{T'_u\}_{u\in V})$ of $G$ are {\em
isomorphic}, and write $\T\simeq\T'$, if there exists an isomorphism $\varphi$
between $T$ and $T'$ that induces an isomorphism between $T_u$ and $T'_u$ for
all $u\in V$, namely $\varphi\big(V(T_u)\big)=V(T_u')$.

A tree model $\T=(T,\{T_u\}_{u\in V})$ of $G$ is {\em minimal} if $|V(T)|$ is
smallest possible among all tree models~of~$G$.
A {\em clique tree} of $G$ is a tree $T$ whose nodes are the maximal cliques of
$G$ such that for all $C,C'\in V(T)$, every $C''$ on the path between $C$ and
$C'$ in $T$ satisfies $C''\supseteq C\cap C'$. 
Every clique tree $T$ of $G$ {\em defines} a tree model $\T_T$ of $G$, where $\T_T=(T,\{T_u\}_{u\in
V})$ and $T_u$ is defined as $T\big[\{C\in V(T)~|~u\in C\}\big]$ for all $u\in V$.

\begin{fact}\label{fact:1}
Let $\T=(T,\{T_u\}_{u\in V})$ be a tree model of $G$. Then the following
statements are equivalent.\smallskip
\begin{compactenum}[(i)]
\item $\T$ is a minimal tree model of $G$.
\item $\T\simeq \T_T$ for some clique tree $T$ of $G$.
\item
For all $XY\in E(T)$, contracting $XY$ in
$T$ and all subtrees $T_u$ containing it yields a tree model of $G'\neq G$.
\item
The mapping $\psi$ defined for $X\in V(T)$ as $\psi(X)=\{u\in V~|~X \in V(T_u)\}$
 is a bijection between the vertices of $T$ and the maximal cliques of $G$.
\end{compactenum}
\end{fact}

\begin{proof}
(i)$\Rightarrow$(iii) and (ii)$\Leftrightarrow$(iv) are clear, while (iii)$\Rightarrow$(iv)$\Rightarrow$(i)
follow from the Helly property of subtrees.
\end{proof}

Note that (iv) in the above claim states, in other words, that the set of all
vertices of $G$ whose subtrees contain $X$ is a maximal clique of $G$. In
particular, for any tree model, the set of such vertices is always a clique of
$G$, but it is not always necessarily a maximal clique.  This is only true for
minimal tree models.

It follows from Fact \ref{fact:1}(i)$\Leftrightarrow$(iii) that every tree model
$(T,\{T_u\}_{u\in V})$ of $G$ can be transformed (by contracting some edges of
the host tree and the subtrees) into a minimal tree model $(T',\{T'_u\}_{u\in
V})$.  Notably, as this transformation involves only contracting edges, it
follows that this does not increase the number of leaves both in the host tree
and the subtrees, namely $|\Ll(T')|\leq |\Ll(T)|$ and $|\Ll(T_u)|\leq
|\Ll(T'_u)|$ for all $u\in V$. 

This observation allows us to focus exclusively on minimal tree models.  Namely,
it shows that if there exists a tree model with minimum number of leaves in the
host tree (subtrees), then there also is a minimal tree model with minimum
number of leaves in the host tree (subtrees).  Consequently, in the remainder of
the paper, all tree models are assumed to be minimal tree models unless
otherwise specified.

Furthermore, using Fact \ref{fact:1}(i)$\Leftrightarrow$(ii), we shall view
minimal tree models of $G$ as tree models defined by clique trees of $G$. We
shall switch between the two viewpoints as needed.

\section{Vertex Leafage is NP-complete}\label{sec:vl_is_npc}

In this section, we prove Theorem \ref{thm:1} stating that calculating the
vertex leafage of a chordal graph is NP-complete. We describe a polynomial time
reduction from the problem {\sc NOT-ALL-EQUAL-$k$-SAT} which is well-known to be
NP-complete \cite{garey-johnson}.\medskip

\begin{proofof}{Theorem \ref{thm:1}}
The problem is clearly in NP as one can easily compute in polynomial time the
number of leaves in subtrees of a given tree model. To prove NP-hardness, we
show a reduction from {\sc NOT-ALL-EQUAL-$k$-SAT}.  By standard arguments \cite{schaefer}, we
may assume, without loss of generality, that the instances to this problem
contain no repeated literals and no negated variables. Thus we can phrase the
problem as follows.\smallskip

\noindent\underline{\sc NOT-ALL-EQUAL-$k$-SAT}\smallskip\\
{\em Instance} $\cal I$: a
collection $C_1,C_2,\ldots,C_m$ of $k$-element subsets of
$\{v_1,\ldots,v_n\}$;\\
{\em Solution} to $\cal I$ (if exists): a set $S\subseteq
\{v_1,\ldots,v_n\}$ such that each $j\in\{1\ldots m\}$ satisfies
$C_j\setminus S\neq \emptyset$ and $S\setminus C_j\neq\emptyset$.
\smallskip

In addition, we may assume the following property of any instance $\cal I$.\smallskip

\noindent($\star$)~ There are no distinct indices $i,i^+$ such that
$v_{i^+}\in C_j$ whenever $v_i\in C_j$.\smallskip

\noindent Indeed, 
if there exist $i\neq i^+$ with
$v_{i^+}\in C_j$ whenever $v_i\in C_j$, then we replace $\cal I$ by another
instance $\cal I^+$ constructed from $\cal I$ by removing $v_i$ and all clauses $C_j$ that contain $v_i$.
If there is a solution to $\cal I$, then clearly $S\setminus\{v_i\}$ is a
solution to $\cal I^+$.  Conversely, if $S$ is a solution to $\cal I^+$, then
either $S$ is a solution to $\cal I$ if $v_{i^+}\in S$, or
$S\cup\{v_i\}$ is a solution to $\cal I$ if otherwise.
\smallskip

Now, for the reduction, we consider an instance $\cal I$ satisfying ($\star$)
and construct a graph, denoted by $G_{\cal I}$, as follows:\smallskip

\begin{compactenum}[(i)]
\item the vertex set of $G_{\cal I}$ consists of $n+m+2$ vertices: $V(G_{\cal
I})=\{v_1,\ldots,v_n,y_1,\ldots,y_m,z_1,z_2\}$,
\item the vertices $\{y_1,\ldots,y_m\}$ form a clique,
\item the vertices $\{v_1,\ldots,v_n,z_1,z_2\}$ form an independent set,
\item each vertex $v_i$ is adjacent to all vertices $y_j$ such that $v_i\in C_j$,
\item the vertices $z_1,z_2$ are adjacent to each vertex of the clique $\{y_1,\ldots,y_m\}$.
\end{compactenum}
\smallskip

\noindent We observe that $G_{\cal I}$ is a split graph with partition into clique
$\{y_1,\ldots,y_m\}$ and independent set $\{v_1,\ldots,$ $v_n,z_1,z_2\}$.

We prove that the vertex leafage of $G_{\cal I}$ is:\smallskip
\begin{compactenum}[(a)]
\item at most $k+1$, and
\item is at most $k$ if and only if there is a solution to $\cal I$. 
\end{compactenum}\smallskip

To do this we analyze the cliques of $G_{\cal I}$.
This is easy, since $G_{\cal I}$ is a split graph; all its maximal cliques are
formed by taking a vertex of the independent set with its neighbourhood.
In particular, the maximal cliques of $G_{\cal I}$ are 
$A=\{z_1,y_1,\ldots,y_m\}$, $B=\{z_2,y_1,\ldots,y_m\}$, and $Q_i=\{v_i\}\cup\{y_j~|~v_i\in
C_j\}$ for each $i\in\{1\ldots n\}$.

\medskip

We first prove (a).  Recall that $\{A,B,Q_1,\ldots,Q_n\}$ is the set of all
maximal cliques of $G_{\cal I}$, and hence, the vertex set of every clique tree
of $G_{\cal I}$.  Each of the vertices $z_1$, $z_2$, and $v_i$, for
$i\in\{1\ldots n\}$, belongs to exactly one of these cliques, namely $A$, $B$,
and $Q_i$, respectively. Also, each $y_j$, for $j\in\{1\ldots m\}$, belongs to
exactly $k+2$ cliques, namely $A$, $B$, and $\{Q_{i_1},\ldots,Q_{i_k}\}$ where
$C_j=\{v_{i_1},\ldots,v_{i_k}\}$. So, as $k\geq 3$, every tree spanning these
cliques has at most $k+1$ leaves. We thus conclude that in every clique tree of
$G_{\cal I}$, each subtree corresponding to a vertex of $G_{\cal I}$ has at most
$k+1$ leaves.  In other words, any clique tree of $G_{\cal I}$ certifies that
$\vl(G_{\cal I})\leq k+1$ which proves (a).

We now prove (b). Let $S$ be a solution to $\cal I$. Construct a tree $T$ with
vertex set $\{A,B,Q_1,\ldots,Q_n\}$ and edge set $\{AB\}\cup\{AQ_i~|~v_i\in
S\}\cup\{BQ_i~|~i\not\in S\}$.  Let us verify that $T$ is a clique tree of
$G_{\cal I}$.  Its vertex set is the set of all maximal cliques of $G_{\cal I}$.
For distinct $i,i^+\in\{1\ldots n\}$, the path between $Q_i$ and $Q_{i^+}$
contains $A$ or $B$ or both, and no other vertex. Note that $Q_i\cap
Q_{i^+}\subseteq\{y_1,\ldots,y_m\}=A\cap B$. This verifies the path between
$Q_i$ and $Q_{i^+}$.  Similarly, the path between $Q_i$ and $A$ or $B$
additionally contains only $A$ or $B$ and we have $Q_i\cap A=Q_i\cap B$ which
verifies this path. That exhausts all paths in $T$ and thus confirms that $T$ is
indeed a clique tree of $G_{\cal I}$.

Let $\T_T=\big(T,\{T_v\}_{v\in V(G_{\cal I})}\big)$ be the tree model
corresponding to $T$.  We analyze its subtrees. First, we consider the subtree
$T_{v_i}$ where $i\in\{1\ldots n\}$. As in (a), we observe that the vertex $v_i$
only belongs to one clique of $G_{\cal I}$, namely $Q_i$.  Thus $|V(T_{v_i})|=1$
implying $|\Ll(T_{v_i})|=0$ by our convention.  Similarly, the vertices $z_1$
and $z_2$ each belong to only one clique, $A$ and $B$ respectively, and we have
$|\Ll(T_{z_1})|=|\Ll(T_{z_2})|=0$.  It remains to consider $T_{y_j}$ for
$j\in\{1\ldots m\}$.  The vertex $y_j$ belongs to the cliques $A$, $B$, and $k$
distinct cliques $Q_{i_1},\ldots,Q_{i_k}$ where
$C_j=\{v_{i_1},\ldots,v_{i_k}\}$.  The cliques $Q_{i_1},\ldots,Q_{i_k}$ are
leaves of $T_{y_j}$ as they are leaves of $T$. However, neither $A$ nor $B$ is a
leaf of $T_{y_j}$.  Indeed, since $S$ is a solution to ${\cal I}$, there are
indices $p,r\in\{1\ldots k\}$ such that $v_{i_p}\in S$ and $v_{i_r}\not\in S$.
Hence, by construction, $T$ contains edges $AQ_{i_p}$ and $BQ_{i_r}$. So,
$T_{y_j}$ contains these edges as well as the edge $AB$. Thus both $A$ and $B$
have at least two neighbours in $T_{y_j}$ and are therefore not leaves of
$T_{y_j}$. Consequently, $|\Ll(T_{y_j})|=|\{Q_{i_1},\ldots,Q_{i_k}\}|=k$ which
implies $v\ell(G_{\cal I})\leq k$ as certified by the tree model $\T_T$.

Conversely, suppose that $v\ell(G_{\cal I})\leq k$. Then there exists a clique
tree $T$ of $G_{\cal I}$ such that the corresponding model
$\T_T=\big(T,\{T_v\}_{v\in V(G_{\cal I})}\big)$ satisfies $|\Ll(T_v)|\leq k$ for
all $v\in V(G_{\cal I})$. We analyze the structure of $T$.  First, we observe
that $AB$ must be an edge of $T$. If otherwise, the path between $A$ and $B$ in
$T$ contains some clique $Q_i$, $i\in\{1\ldots n\}$. As $T$ is a clique tree, we
conclude $\{y_1\ldots y_m\}=A\cap B\subseteq Q_i=\{v_i\}\cup\{y_j~|~v_i\in
C_j\}$. But then $v_i$ belongs to each $C_j$, $j\in\{1\ldots m\}$, and since
$n\geq k\geq 3$, this contradicts ($\star$).  Similarly, we show that each
$Q_i$, $i\in\{1\ldots n\}$ is a leaf of $T$. If otherwise, some $Q_i$ has at
least two neighbours in $T$. These cannot be $A,B$ as this would imply a
triangle in $T$, since $AB$ is an edge of $T$.  Thus $Q_i$ is adjacent to
$Q_{i^+}$ for some $i^+\in\{1\ldots n\}$.  As $T$ is a tree, we have that either
$Q_{i^+}$ lies on the path from $A$ to $Q_i$, or $Q_i$ lies on the path from $A$
to $Q_{i^+}$.  By symmetry, we may assume the former.  Thus, since $T$ is a
clique tree, we conclude $\{y_j~|~v_i\in C_j\}=A\cap Q_i\subseteq
Q_{i^+}=\{v_{i^+}\}\cup\{y_j~|~v_{i^+}\in C_j\}$. So $v_{i^+}\in C_j$ whenever
$v_i\in C_j$, contradicting ($\star$).

Now, we are ready to construct a set $S\subseteq\{v_1,\ldots,v_n\}$ as follows:
for each $i\in\{1 \ldots n\}$, we put $v_i$ in $S$ if $AQ_i$ is an edge of $T$.
We show that $S$ is a solution to $\cal I$. If not, there exists $j\in\{1\ldots
m\}$ such that either $S\supseteq C_j$ or $S\cap C_j=\emptyset$.  We look at the
subtree $T_{y_j}$ corresponding to the vertex $y_j$. Recall that $y_j$ belongs
to cliques $A$, $B$, and $k$ cliques $Q_{i_1},\ldots,Q_{i_k}$ where
$C_k=\{v_{i_1},\ldots,v_{i_k}\}$.  The cliques $Q_{i_1},\ldots,Q_{i_k}$ are
leaves of $T_{y_j}$ because they are leaves of $T$ (as proved above).  If
$S\subseteq C_j$, we have, by construction, that $A$ is the unique neighbour of
each of the cliques $Q_{i_1},\ldots,Q_{i_k}$ in $T$. Consequently, none of the
cliques $Q_{i_1},\ldots,Q_{i_k}$ is adjacent to $B$ in $T$.  This shows that $B$
is only adjacent to $A$ in $T_{y_j}$, and hence, is a leaf.  But then
$|\Ll(T_{y_j})|=|\{Q_{i_1},\ldots,Q_{i_k},B\}|=k+1$, contradicting our
assumption about $T$.  Similarly, if $S\cap C_j=\emptyset$, the cliques
$Q_{i_1},\ldots,Q_{i_k}$ are only adjacent to $B$ and not to $A$, in which case,
$A$ is a leaf of $T_{y_j}$ leading~to~the~same~contradiction.

Therefore, $S$ must indeed be a solution to $\cal I$ and that concludes the proof.
\end{proofof}
\vspace{-2ex}

\section{Vertex Leafage Parameterized by Leafage} \label{sec:fixed_l}

In this section, we discuss calculating vertex leafage in chordal graphs of
bounded leafage.  Namely, we prove Theorem \ref{thm:2}, that is, for a fixed
$\ell$, we demonstrate how to calculate the vertex leafage of an $n$-vertex
chordal graph $G$ with $\ell(G)\leq \ell$ in polynomial time, namely, in time
$n^{O(\ell)}$. We do this by enumerating clique trees of $G$ with respect to
high ($\geq 3$) degree nodes.  The enumeration is based on the observation that
the number of high-degree nodes in a tree is directly related to the number of
leaves.  This goes as follows.

For a tree $T$, let $\Hh(T)$ denote the set of nodes of $T$ of degree $\geq 3$,
and let $\Ee(T)$ denote the set of edges of $T$ incident to the nodes in
$\Hh(T)$.  Further, let $n_i$ denoted the number of nodes of degree $i$ in $T$.
Then\medskip

\noindent($\star$)
\hfill
$|\Hh(T)|=\Sum_{i\geq 3} n_i\leq |\Ee(T)|\leq \Sum_{i\geq 3} (i-2)n_i =
2|E(T)|-2|V(T)| + n_1 = |\Ll(T)|-2$
\hfill\mbox{}
\smallskip

\noindent In particular, if $|\Ll(T)|$ is bounded, then so is $|\Hh(T)|$ and
$|\Ee(T)|$. This will become useful later.

Recall that the vertex set of every clique tree of $G$ is the set of all maximal
cliques of $G$. Notably, all clique trees have the same vertex set.  Let $\C(G)$
denote the clique graph of $G$, i.e., the graph whose nodes are the maximal
cliques of $G$ and where two nodes are adjacent if and only if the corresponding
maximal cliques intersect.  It is well-known \cite{gavril-gener,shibata} that
every clique tree of $G$ is a spanning tree of $\C(G)$.

Our algorithm is based on the following lemma.

\begin{lemma}\label{lem:alg1}
There is an $O(n^3)$ time algorithm that, given an $n$-vertex chordal graph $G$
and a set $F\subseteq E(\C(G))$, decides if there exists a clique tree $T$ of
$G$ with $\Ee(T)=F$ and constructs such a tree if one exists.
\end{lemma}

\begin{proof}
We describe an algorithm for the problem as follows.

\DontPrintSemicolon

\begin{algorithm}[H]
\caption{\label{alg:1}}
\vskip -2ex
\KwIn{A chordal graph $G$ and a set $F\subseteq E(\C(G))$.}
\KwOut{A clique tree $T$ of $G$ with $\Ee(T)=F$, or report that no such tree exists.}
\smallskip

Construct a graph $G'$ as follows:\hspace{200em}
\mbox{}\qquad $V(G')=V(G)\cup\{v_e~|~e\in F\}$\hspace{200em}
\mbox{}\qquad $E(G')=E(G)\cup\{uv_e~|~e=CC', u\in C\cup C'\}\cup
\{v_ev_{e'}~|~e\cap e'\neq\emptyset\}$\;

\If{$G'$ is chordal}
{
Construct a clique tree $T'$ of $G'$ with minimum number of leaves.\;

Construct a tree $T$ from $T'$ by renaming each node $C'\in V(T')$ to $C'\cap
V(G)$\;

\If{$T$ is a clique tree of $G$ and $\Ee(T)=F$}{
\Return{$T$}\;
}
}
\Return{``no such tree exists''}\;
\vskip -2ex
\end{algorithm}

We now prove correctness of the above algorithm.  For simplicity, we shall refer
to any clique tree $T$ with $\Ee(T)=F$ as a ``solution''.
First, observe that if the algorithm returns the tree $T$ in line 6, then this
is indeed a solution. This proves that if there is no solution, the algorithm
provides the correct answer in~line~7.

Thus, for the rest of the proof, we may assume that a solution exists. Namely we
shall assume there is a clique tree $T^*$ of $G$ satisfying $\Ee(T^*)=F$.
For every maximal clique $C$ of $G$, define $\varphi(C)=C\cup\{v_e~|~C\in e\}$.

In the following claim, we discuss the properties of the graph $G'$ constructed
in line 1.

\begin{claim}\label{clm:a1}
$G'$ is chordal, satisfies $\ell(G')\leq |\Ll(T^*)|$, and $\varphi$ is a
bijection between the maximal cliques of $G$ and $G'$.
\end{claim}

\begin{proofclaim}
To prove the claim, we construct a minimal tree model of $G'$ as follows.  Let
$\T_{T^*}=(T^*,\{T^*_u\}_{u\in V(G)})$ be the minimal tree model of $G$ that is
defined by the clique tree $T^*$, namely $T^*_u=T[\{C\in V(T^*)~|~u\in C\}]$.
For each edge $e=CC'\in F$, define $T^*_{v_e}=T^*[\{C,C'\}]$.  Finally, let
$\T^+=(T^*,\{T^*_u\}_{u\in V(G)}\cup\{T^*_{v_e}\}_{e\in F}\})$.

It is easy to verify that $\T^+$ is a tree model of $G'$. In particular, each
subtree in the collection is a connected subgraph of $T^*$. This follows from
the fact that $T^*$ is a clique tree and that $F=\Ee(T^*)\subseteq E(T^*)$.
Further, for each edge $e=CC'$ in $F$, we see that the subtree $T^*_{v_e}$
intersects only subtrees $T^*_u$ where $C$ or $C'$ is in $V(T^*_u)$, i.e., those
where $u\in C\cup C'$.  Moreover, $T^*_{v_e}$ only intersects subtrees
$T^*_{v_{e'}}$ where $C$ or $C'$ is in $V(T^*_{v_{e'}})$, i.e., those where
$e\cap e'\neq\emptyset$.  This corresponds precisely to the definition of $G'$.

Thus, we conclude that $G'$ is indeed a chordal graph, and $\ell(G')\leq
|\Ll(T^*)|$ as $\T^+$ is a particular tree model of $G'$ and $T^*$ is its host
tree.  Morever, we see that $\T^+$ is actually a minimal tree model of $G'$.
Indeed, if there were a tree model of $G'$ with less than $|V(T^*)|$ nodes in
its host tree, then by removing subtrees corresponding to the vertices $\{v_e
~|~e\in F\}$ we would obtain a tree model of $G$ whose host tree has less than
$|V(T^*)|$ nodes. But this would contradict the minimality of $\T_{T^*}$.

This implies, by Fact \ref{fact:1}(ii), that there exists a clique tree $T^+$ of
$G'$ that defines $\T^+$, i.e., $\T^+=\T_{T^+}$. Namely, there is an isomorphism
between $T^+$ and the host tree $T^*$ of $\T^+$ where each node $C\in V(T^*)$
corresponds to the set of all vertices of $G'$ whose subtrees contain $C$, i.e.,
the set $\{u\in V(G)~|~C\in V(T_u)\}\cup\{v_e~|~C\in V(T_{v_e})\}$ which is
exactly $\varphi(C)$.  In other words, $V(T^+)=\{\varphi(C)~|~C\in V(T^*)\}$,
and consequently, $\varphi$~constitutes an isomorphism between $T^*$ and $T^+$.
As one is a clique tree of $G$ and the other a clique tree of $G'$, we conclude
that $\varphi$ is a bijection between the maximal cliques of $G$ and $G'$.
\end{proofclaim}

This proves that the test in Line 2 succeds.  Now, consider the trees $T'$ and
$T$ constructed in line 3 and 4. Notably, $T'$ is a clique tree of $G'$ with
$|\Ll(T')|=\ell(G')$.

\begin{claim}\label{clm:a2}
$T$ is a clique tree of $G$.
\end{claim}

\begin{proofclaim}
Recall that $T$ is obtained from $T'$ by renaming each node $C'$ of $T'$ to
$C'\cap V(G)$.  Moreover, by (\ref{clm:a1}), the mapping $\varphi$ is a
bijection between the maximal cliques of $G$ and $G'$. Namely, for each $C'\in
V(T)$, the set $C=\varphi^{-1}(C')$ is a maximal clique of $G$. Therefore, we
can write \smallskip

\noindent\hfill
$C'\cap V(G) = \varphi(C)\cap V(G) = \big(C\cup\{v_e~|~C\in e\}\big)\cap V(G) =
C =\varphi^{-1}(C')$.
\hfill\mbox{}
\smallskip

\noindent This proves that the vertex set of $T$ is precisely the set of maximal
cliques of $G$, and $\varphi$ is an isomorphism between $T$ and $T'$, by the
construction of $T$.  To see that $T$ is indeed a clique tree of $G$, it remains
to prove the ``connectivity condition'' for $T$.  Namely, consider nodes
$C_1,C_2\in V(T)$ and a node $C_3$ on the path in $T$ between $C_1$ and $C_2$.
Since $\varphi$ is an isomorphism between $T$ and $T'$, we have $\varphi(C_i)\in
V(T')$ for $i=1,2,3$ and $\varphi(C_3)$ lies on the path in $T'$ between
$\varphi(C_1)$ and $\varphi(C_2)$.  Thus, we conclude
$\varphi(C_3)\supseteq\varphi(C_1)\cap\varphi(C_2)$ because $T'$ is a clique
tree.  So we write $C_3 = \varphi(C_3)\cap V(G) \supseteq \varphi(C_1)\cap
\varphi(C_2)\cap V(G) = C_1\cap C_2$.
\end{proofclaim}

This proves that $T$ is a clique tree of $G$. Notably, as $T^*$ is also a clique
tree of $G$, we have that both $T$ and $T^*$ have the same vertex set, i.e.,
$V(T)=V(T^*)$.  We now look at the edges of $T$.

\begin{claim}\label{clm:a3}
$F\subseteq E(T)$
\end{claim}

\begin{proofclaim}
Consider an edge $e=CC'\in F$, and recall the definition of $\varphi$ and the
claim (\ref{clm:a1}). From this it follows that $\varphi(C)$ and $\varphi(C')$
are the only maximal cliques of $G'$ that contain $v_e$. As $\varphi(C)$ and
$\varphi(C')$ are also nodes of $T'$ which is a clique tree of $G'$, we conclude
that every maximal clique on the path in $T'$ between $\varphi(C)$ and
$\varphi(C')$
also contains $v_e$.  But, as $v_e$ is in no other maximal clique of $G'$,
this is only possible if $\varphi(C)$ and $\varphi(C')$ are adjacent in $T'$.
Consequently, $C$ and $C'$ are adjacent in $T$, namely $e\in E(T)$.
\end{proofclaim}

\begin{claim}\label{clm:a4}
$\Hh(T^*)\subseteq \Hh(T)$ and each $C\in\Hh(T^*)$ satisfies
$N_{T^*}(C)\subseteq N_T(C)$.
\end{claim}

\begin{proofclaim}
Consider $C\in\Hh(T^*)$, namely $C$ is a node of $T^*$ with at least
three neighbours in $T^*$. Then, by the
definition of $\Ee(T^*)$, all edges incident to $C$ in $T^*$ belong to
$\Ee(T^*)$. As $\Ee(T^*)=F$ and $F\subseteq E(T)$ by (\ref{clm:a3}), the edges
incident to $C$ in $T^*$ are also edges of $T$.  In other words, every neighbour
of $C$ in $T^*$ is a neighbour of $C$ in $T$, namely $N_T(C)\supseteq
N_{T^*}(C)$. Thus $C$ has at least three neighbours in $T$ implying
$C\in\Hh(T)$.
\end{proofclaim}

\begin{claim}\label{clm:a5}
$\Hh(T)=\Hh(T^*)$ and $\Ee(T)=\Ee(T^*)$.
\end{claim}

\begin{proofclaim}
By (\ref{clm:a3}), we conclude $\Hh(T)\supseteq \Hh(T^*)$.  For the converse, we
calculate using (\ref{clm:a1}) and $(\star)$ as follows.
\medskip

$\ell(G')\leq |\Ll(T^*)|=2+\Sum_{C\in\Hh(T^*)}\big(\deg_{T^*}(C)-2\big)\leq 
2+\Sum_{C\in\Hh(T)}\big(\deg_{T}(C)-2)=|\Ll(T)|=\ell(G')$
\smallskip

\noindent Note that the second inequality follows from (\ref{clm:a4}) and the
fact that $\deg_{T}(C)\geq 3$ for all $C\in\Hh(T)$, while the last equality is
by $\ell(G')=|\Ll(T')|$ and the fact that $T$ and $T'$ is isomorphic.

It follows that the inequalities in the above formula are, in fact, equalities.
Therefore, using (\ref{clm:a4}), we conclude that $\Hh(T)=\Hh(T^*)$ and every
$C\in \Hh(T^*)$ satisfies $N_T(C)=N_{T^*}(C)$. To see this, recall that each
$C\in\Hh(T^*)$ contributes to the sum on the right at least as much as to the
sum on the left, since $N_T(C)\supseteq N_{T^*}(C)$ by (\ref{clm:a4}).  Further,
every $C\in \Hh(T)$ has a positive contribution to the sum on the right as
$\deg_T(C)\geq 3$ by the definition of $\Hh(T)$.  Thus, since the two sums are
equal, the only possibility is that $\Hh(T)=\Hh(T^*)$ and that each $C\in
\Hh(T^*)$ satisfies $N_T(C)=N_{T^*}(C)$ as claimed.

To conclude the proof, recall that $\Ee(T)$, resp. $\Ee(T^*)$, is the set of
edges of $T$, resp. $T^*$, incident to the nodes in $\Hh(T)$, resp. $\Hh(T^*)$.
As $\Hh(T)=\Hh(T^*)$ and each $C\in \Hh(T)=\Hh(T^*)$ is incident to the same set
of edges in $T$ and $T^*$ for it satisfies $N_T(C)=N_{T^*}(C)$, we conclude that
$\Ee(T)=\Ee(T^*)$.
\end{proofclaim}

This and (\ref{clm:a2}) prove that $T$ is indeed a solution, namely $T$ is a
clique tree of $G$ with $\Ee(T)=\Ee(T^*)=F$. Hence, the test in Line 5 succeds
and the algorithm correctly return a solution in Line 6.

That concludes the proof of correctness of the algorithm. To address the
complexity, let $n=|V(G)|$  as usual. First, we note that we may assume that $F$
contains at most $n-1$ edges as no clique tree of $G$ has more than $n$
edges. If this is not so, we can safely report that no solution exists. Thus, as
$G'$ has $|V(G)|+|F|=O(n)$ vertices, we conclude that step 3 takes $O(n^3)$ time
using the algorithm of \cite{leafage}. All other steps clearly take at most
$O(n^2)$ time. Notably, in step 2 we use a linear time algorithm from
\cite{tarjan}.

Thus the total complexity is $O(n^3)$ as promised. That concludes the proof.
\end{proof}

Finally, we are ready to prove Theorem \ref{thm:2}.
\medskip

\begin{proofof}{Theorem \ref{thm:2}}
Let $G$ be a chordal graph with $\ell(G)\leq \ell$.  By Corollary \ref{cor:4}
(proven in \S \ref{sec:fixed_vl}),
there exists a tree model of $G$ that simultaneously minimizes both the leafage
and the vertex leafage.  By the remarks in \S \ref{sec:clique-tree}, there
is also a clique tree of $G$ with this property; let $T^*$ denote this clique
tree.  Note that $|\Ll(T^*)|=\ell(G)$ and $|\Ll(T^*_u)|\leq \vl(G)$ for all
$u\in V(G)$ where $T^*_u=T^*\big[\{C\in V(T^*)~|~u\in C\}\big]$.

We show that it suffices to know the set $\Ee(T^*)$ to find a tree model that
minimizes the vertex leafage.

\begin{claim}\label{clm:b1}
If $T$ is a clique tree of $G$ with $\Ee(T)=\Ee(T^*)$, then $T$ minimizes the
vertex leafage.
\end{claim}

\begin{proofclaim}
Consider $u\in V(G)$.  We need to show that $|\Ll(T_u)|\leq \vl(G)$  where
$T_u=T\big[\{C\in V(T)~|~u\in C\}\big]$.  

First, we observe that $\Ee(T)=\Ee(T^*)$ implies $\Hh(T)=\Hh(T^*)$ and each
$C\in \Hh(T)=\Hh(T^*)$ has the same neighbourhood in both $T$ and $T^*$, i.e.,
$N_T(C)=N_{T^*}(C)$.
Next, we remark that if a node has degree~$\geq 3$ in $T_u$, then it also has
degree $\geq 3$ in $T$, since $T_u$ is an subgraph of $T$. In other words, we have
$\Hh(T_u)\subseteq\Hh(T)$. Further, we observe that each $C\in V(T_u)$
satisfies $N_{T_u}(C)=N_{T}(C)\cap V(T_u)$, since $T_u$ is an induced subgraph
of $T$.  By the same token, $N_{T^*_u}(C)=N_{T^*}(C)\cap V(T^*_u)$
for each $C\in V(T^*_u)$.  Finally, we note that $V(T_u)=V(T^*_u)$, since
$V(T)=V(T^*)$.  Thus, for each $C\in \Hh(T_u)$, we can write
\smallskip

\noindent\mbox{}\hfill
$N_{T_u}(C)=N_T(C)\cap V(T_u)=N_{T^*}(C)\cap V(T^*_u)=N_{T^*_u}(C)$.
\hfill\mbox{}
\smallskip

\noindent This implies $C\in\Hh(T^*_u)$ and $\deg_{T_u}(C)=\deg_{T^*_u}(C)$ for
all $C\in \Hh(T_u)$.  Thus, we calculate by $(\star)$.\medskip

\noindent\mbox{}\hfill
$|\Ll(T_u)|=2+\Sum_{C\in \Hh(T_u)} \big(\deg_{T_u}(C)-2\big) \leq 2+\Sum_{C\in
\Hh(T^*_u)}\big(\deg_{T^*_u}(C)-2\big) = |\Ll(T^*_u)|\leq \vl(G)$
\hfill\mbox{}
\smallskip

\noindent For the inequality to hold, also note that 
$\deg_{T^*_u}(C)\geq 3$ for each $C\in \Hh(T^*_u)$, by definition.
\end{proofclaim}

This claim allows us to finally formulate our algorithm.  We need to introduce
additional of notation.  Let $F\subseteq E(\C(G))$.  If there exists a clique tree
$T$ with $\Ee(T)=F$, then define $\vl_F=\max_{u\in V(G)}|\Ll(T_u)|$
where $T_u=T\big[\{C\in V(T)~|~u\in C\}\big]$. If such a tree does not exist,
define $\vl_F=+\infty$. Observe that $\vl_{\Ee(T^*)}\leq\vl(G)$.

Our algorithm tries all possible sets $F\subseteq\C(G)$ of size at most $\ell-2$ as
candidates for $\Ee(T^*)$ and chooses one that that minimizes $\vl_F$.  If
$F_{opt}$ is this set, the algorithm outputs a clique tree $T_{opt}$ of $G$
with $\Ee(T_{opt})=F_{opt}$.

We claim that this algorithm correctly finds a clique tree of $G$ that minimizes
the vertex leafage.  By~$(\star)$, we observe that $\Ee(T^*)\leq
|\Ll(T^*)|-2\leq\ell-2$.  Thus, the algorithm must, at some point, consider as $F$
the set $\Ee(T^*)$. For this $F$, we have $\vl_F=\vl_{\Ee(T^*)}\leq\vl(G)$.  By the
minimality of $F_{opt}$, we conclude $\vl_{F_{opt}}\leq
\vl_{\Ee(T^*)}\leq\vl(G)$. Hence, $\vl_{F_{opt}}<\infty$ and so the tree
$T_{opt}$ exists.  Moreover, $\vl_F\geq\vl(G)$ for all sets~$F$, by the
definition of $\vl(G)$.  Thus, we conclude $\vl_{F_{opt}}=\vl(G)$ and
consequently by (\ref{clm:b1}), $T_{opt}$ is a clique tree of $G$ that minimizes
the vertex leafage.  This proves the correctness of the algorithm.

Finally, let us analyze the complexity. Let $n=|V(G)|$ as usual.
Recall that $G$ has at most $n$ maximal cliques. Thus there are at most $n^2$
edges in $\C(G)$, and hence, at most $n^{2\ell-4}$ choices for the set $F$.
For each choice of $F$, we use Lemma \ref{lem:alg1} to find a clique tree $T$
with $\Ee(T)=F$ if it exists. This takes $O(n^3)$ for each $F$, including the
calculation of $\vl_F$. This yields, altogether, running time
$O(n^{2\ell-1})=n^{O(\ell)}$ as promised.
\end{proofof}

We have shown how to calculate vertex leafage in polynomial time when the input
graph has bounded leafage. It remains open whether this problem is fixed
parameter tractable (FPT) with respect to leafage.

\section{Vertex Leafage with Optimum Leafage} \label{sec:fixed_vl}

In this section, we prove Theorem \ref{thm:3} and Corollary \ref{cor:4}. Namely,
we demonstrate that the algorithm from \cite{esa}, solving the leafage problem,
satisfies the claim of Theorem \ref{thm:3}. This algorithm, given a chordal
graph $G$, outputs a clique tree of $G$ with minimum possible number of leaves.
This is done by starting from an arbitrary clique tree $T$ of $G$, and
iteratively decreasing the number of leaves of $T$ as long as possible.

We observe (and formally prove later in this section) that this process has the additional property that it never increases
the number of leaves in the subtrees of the tree model $\T_T$ defined by $T$. In
other words, if $T^*$ is the clique tree resulting from this process, then
$\T^*=\T_{T^*}$ satisfies the claim of Theorem \ref{thm:3}.  This will imply
that if the starting clique tree $T$ realizes the vertex leafage of $G$, then
$\T^*=\T_{T^*}$ satisfies the claim of Corollary \ref{cor:4}.

For the proof of the above, we need to explain the inner workings of the
algorithm from \cite{esa}.  This algorithm, in place of clique trees, operates
on the so-called token assignments defined as follows.

For a chordal graph $G$, a {\em token assignment} of $G$ is a function $\tau$ that
assigns to every maximal clique $C$ of $G$, a multiset $\tau(C)$ of subsets of
$C$.  We use the word {\em token} for the members of $\tau(C)$. Note that the
same subset may appear in $\tau(C)$ many times.  We focus on special token
assignment that arise from clique trees.

The token assignment {\em defined} by a clique tree $T$ of $G$, and denoted by
$\varepsilon_T$, assigns to every maximal clique $C$ of $G$, the multiset
$\varepsilon_T(C)=\{C\cap C'~|~CC'\in E(T)\}$.  In other words,
$\varepsilon_T(C)$ consists of the intersections of~$C$ with its neighbours in
$T$.  A token assignment $\tau$ is {\em realizable} if there is a clique tree
$T$ of $G$ such that $\tau=\varepsilon_T$.

\begin{figure}[h]
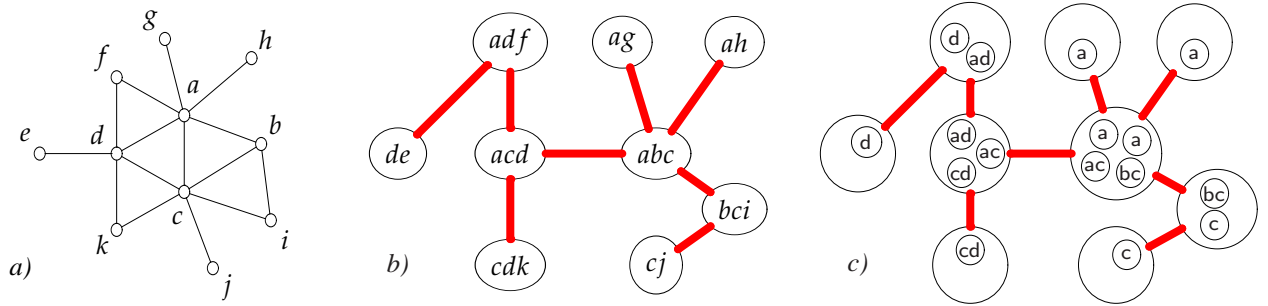

\centering
$
\xy/r3pc/:
(0.2,0)*[o][F]{\phantom{s}}="a";
(1,0.8)*[o][F]{\phantom{s}}="b";
(1,0)*[o][F]{\phantom{s}}="c";
(1,-0.8)*[o][F]{\phantom{s}}="d";
(1.7,.4)*[o][F]{\phantom{s}}="e";
(1.7,-.4)*[o][F]{\phantom{s}}="f";
(2.4,1)*[o][F]{\phantom{s}}="g";
(1.5,1.2)*[o][F]{\phantom{s}}="g'";
(2.5,.1)*[o][F]{\phantom{s}}="h";
(2.6,-.7)*[o][F]{\phantom{s}}="i";
(2,-1.2)*[o][F]{\phantom{s}}="j";
{\ar@{-} "a";"c"};
{\ar@{-} "b";"c"};
{\ar@{-} "c";"d"};
{\ar@{-} "b";"e"};
{\ar@{-} "c";"e"};
{\ar@{-} "c";"f"};
{\ar@{-} "d";"f"};
{\ar@{-} "e";"f"};
{\ar@{-} "e";"g"};
{\ar@{-} "e";"g'"};
{\ar@{-} "e";"h"};
{\ar@{-} "f";"h"};
{\ar@{-} "f";"i"};
{\ar@{-} "f";"j"};
{\ar@{-} "h";"i"};
"c"+(-0.2,0.2)*{d};
"e"+(0.1,0.3)*{a};
"f"+(-0.07,-0.27)*{c};
"h"+(0.15,0.2)*{b};
"a"+(-0.15,0.2)*{{e}};
"b"+(-0.15,0.2)*{{f}};
"d"+(-0.15,-0.15)*{{k}};
"g"+(0.15,0.2)*{{h}};
"i"+(0.15,-0.2)*{{i}};
"j"+(0.15,-0.2)*{{j}};
"g'"+(-0.15,0.2)*{{g}};
(0,-1.2)*{\mbox{\em a)}};
\endxy
\quad\qquad
\xy/r3.5pc/:
(0,0)*++[o][F]{de}="a";
(1,1)*++[o][F]{adf}="b";
(1,0)*++[o][F]{acd}="c";
(1,-1)*++[o][F]{cdk}="d";
(2,1)*++[o][F]{ag}="e";
(3,1)*++[o][F]{ah}="e'";
(2.3,0)*++[o][F]{abc}="f";
(2.3,-1)*++[o][F]{cj}="g";
(3,-0.5)*++[o][F]{bci}="h";
{\ar@{-}@*{[red]}@*{[|<3pt>]} "a";"b"};
{\ar@{-}@*{[red]}@*{[|<3pt>]} "b";"c"};
{\ar@{-}@*{[red]}@*{[|<3pt>]} "c";"d"};
{\ar@{-}@*{[red]}@*{[|<3pt>]} "f";"e"};
{\ar@{-}@*{[red]}@*{[|<3pt>]} "f";"e'"};
{\ar@{-}@*{[red]}@*{[|<3pt>]} "c";"f"};
{\ar@{-}@*{[red]}@*{[|<3pt>]} "h";"g"};
{\ar@{-}@*{[red]}@*{[|<3pt>]} "f";"h"};
(0,-1)*{\mbox{\em b)}};
\endxy
\qquad
\xy/r3.5pc/:
(0,0)*++++[o][F]{\phantom{s}}="a";
"a"*{\xy/2pc/:
(0,0)*{};
(0.5,-0.5)*+[o][F]{\phantom{\bf _d}}="z";
"z"*{\sf _{d}};
\endxy};
(1,1)*+++++[o][F]{}="b";
"b"*{\xy/2pc/:
(0.4,0)*{};
(0,-0.5)*{};
(0,0.2)*+[o][F]{\phantom{\bf _4}}="z";
"z"*{\sf _{d}};
(-0.3,-0.2)*+[o][F]{\phantom{\bf _4}}="y";
"y"*{\sf _{ad}};
\endxy};
(1,0)*+++++[o][F]{}="c";
"c"*{\xy/2pc/:
(0,0.5)*{};
(-0.3,0.1)*+[o][F]{\phantom{\bf _4}}="z";
"z"*{\sf _{cd}};
(0,-0.35)*+[o][F]{\phantom{\bf _4}}="x";
"x"*{\sf _{ac}};
(0.3,0.1)*+[o][F]{\phantom{\bf _4}}="y";
"y"*{\sf _{ad}};
\endxy};
(1,-1)*++++[o][F]{\phantom{s}}="d";
"d"*{\xy/2pc/:
(0,0)*{};
(0.6,0)*+[o][F]{\phantom{\bf _4}}="y";
"y"*{\sf _{cd}};
\endxy};
(3,1)*++++[o][F]{\phantom{s}}="e'";
"e'"*{\xy/2pc/:
(0,0)*{};
(-0.6,0)*+[o][F]{\phantom{\bf _4}}="z";
"z"*{\sf _{a}};
\endxy};
(2,1)*++++[o][F]{\phantom{s}}="e";
"e"*{\xy/2pc/:
(0,0)*{};
(-0.6,0)*+[o][F]{\phantom{\bf _4}}="z";
"z"*{\sf _{a}};
\endxy};
(2.3,0)*+++++[o][F]{\phantom{s}}="f";
"f"*{\xy/2pc/:
(-0.5,0)*{};
(-0.2,0.35)*+[o][F]{\phantom{\bf _4}}="z";
"z"*{\sf _{ac}};
(0.2,-0.3)*+[o][F]{\phantom{\bf _4}}="x";
"x"*{\sf _{a}};
(0.3,0.22)*+[o][F]{\phantom{\bf _4}}="y";
"y"*{\sf _{a}};
(-0.3,-0.2)*+[o][F]{\phantom{\bf _4}}="w";
"w"*{\sf _{bc}};
\endxy};
(2.3,-1)*++++[o][F]{\phantom{s}}="g";
"g"*{\xy/2pc/:
(0,0)*{};
(0.45,-0.55)*+[o][F]{\phantom{\bf _4}}="y";
"y"*{\sf _{c}};
\endxy};
(3.2,-0.5)*+++++[o][F]{}="h";
"h"*{\xy/2pc/:
(0.4,0)*{};
(0,-0.5)*{};
(0.25,-0.2)*+[o][F]{\phantom{\bf _4}}="z";
"z"*{\sf _{bc}};
(-0.25,-0.2)*+[o][F]{\phantom{\bf _4}}="y";
"y"*{\sf _{c}};
\endxy};
{\ar@{-}@*{[red]}@*{[|<3pt>]} "a";"b"};
{\ar@{-}@*{[red]}@*{[|<3pt>]} "b";"c"};
{\ar@{-}@*{[red]}@*{[|<3pt>]} "c";"d"};
{\ar@{-}@*{[red]}@*{[|<3pt>]} "f";"e"};
{\ar@{-}@*{[red]}@*{[|<3pt>]} "f";"e'"};
{\ar@{-}@*{[red]}@*{[|<3pt>]} "c";"f"};
{\ar@{-}@*{[red]}@*{[|<3pt>]} "h";"g"};
{\ar@{-}@*{[red]}@*{[|<3pt>]} "f";"h"};
(0,-1)*{\mbox{\em c)}};
\endxy$
\caption{{\em a)} Example chordal graph $G$, {\em b)} clique tree $T$ of $G$,
{\em c)} token assignment $\tau = \varepsilon_T$.\label{fig:token-ass}}
\end{figure}

(See Figure \ref{fig:token-ass} for an illustration of these concepts.)

Notice that the token assignment $\tau=\varepsilon_T$ contains all the
information needed to determine the number of leaves in $T$ and also the number
of leaves in the subtrees of the corresponding model $\T_T$. We summarize this as follows.

\begin{lemma}\label{lem:c1}
Let $G$ be a chordal graph, let $T$ be a clique tree of $G$, and let
$\T_T=\big(T,\{T_u\}_{u\in V(G)}\big)$ denote the tree model of $G$ defined by
$T$.  Let $\tau=\varepsilon_T$, and define $\tau_u(C)=\{S~|~S\in\tau(C),u\in
S\}$ for each $u\in V(G)$.  Then\medskip

\begin{compactitem}
\item $\deg_T(C)=|\tau(C)|$ for all $C\in V(T)$, and
\item $\deg_{T_u}(C)=|\tau_u(C)|$ for all $u\in V(G)$ and all $C\in V(T_u)$.
\end{compactitem}

\noindent Consequently, $\Ll(T)=\Big\{C~\Big|~|\tau(C)|=1\Big\}$ and
$\Ll(T_u)=\Big\{C~\Big|~|\tau_u(C)|=1\Big\}$ for all $u\in V(G)$.
\end{lemma}

\noindent
In particular, while there can be multiple clique trees defining the same token
assignment, these clique trees will have the same sets of leaves and
consequently we do not need to distinguish them from one another.  In other
words, it suffices to maintain that the token assignment we consider corresponds
to some clique tree of $G$. This can be tested easily by applying four
particular conditions as described in \cite{esa}. As we do not use this test here
directly, we omit further details. (For more, see  \cite[Theorem 6]{esa}.)
\smallskip

Now, we are finally ready to explain the main steps of the algorithm from
\cite{esa}.  The algorithm is given a chordal graph $G$ and a clique tree $T$ of
$G$.  It starts by constructing the token assignment $\tau=\varepsilon_{T}$.
Then it proceeds iteratively.  During each iteration step, a current token
assignment $\tau$ is examined to determine if there exists a different token
assignment corresponding to a clique tree with fewer leaves.  This is done by
checking for an {\em augmenting path}  in $\tau$, which is a specific sequence
of {\em token moves} (see definitions below). If an augmenting path exists, we
pick the shortest such path and exchange tokens along the path. This results in
a new token assignment $\tau$ that corresponds to a clique tree with fewer
leaves. If no augmenting path exists, we arrive at an optimal solution (i.e., a token assignment whose corresponding clique trees all have $\ell(G)$ leaves) and we
output this solution.  We summarize the above procedure as Algorithm
\ref{alg:leafage}. Below we provide the missing definitions.

Let $G$ be a chordal graph and $\tau$ be a token assignment of $G$.  A {\em
token move} is an ordered triple ($C_1$, $C_2$, $S$) where $C_1,C_2$ are maximal
cliques of $G$ and $S\in\tau(C_1)$.  For a token move $(C_1,C_2,S)$, we write
$\tau\div (C_1,C_2,S)$ to denote the token assignment $\tau'$ that is the result
of moving $S$ from $\tau(C_1)$ to $\tau(C_2)$.  Namely\footnote{Note that as
both $\tau(C_1)$ and $\tau'(C_1)$ are multisets, to obtain $\tau'(C_1)$ we only
remove one instance of $S$ from $\tau(C_1)$ in case $S$ appears in $\tau(C_1)$
several times. This is consistent with the semantics of the set difference for
multisets.}, we have $\tau'(C_1)=\tau(C_1)\setminus\{S\}$ and
$\tau'(C_2)=\tau(C_2)\cup\{S\}$, while $\tau'(C)=\tau(C)$ for all other
$C\not\in\{C_1,C_2\}$.

A sequence of token moves $(C_1, C_2, S_1)$, $(C_2, C_3, S_2)$, $\ldots$,
$(C_{k-1}, C_k, S_{k-1})$ where $k>1$ is an {\em augmenting path} of $\tau$ if
$|\tau(C_k)|=1$ and each $j\in\{1\ldots k-1\}$ satisfies

(i)~$\tau\div(C_j, C_{j+1}, S_j)$ is a realizable token
assignment\footnote{i.e., it corresponds to a clique tree of $G$.},
and \quad (ii)~ $|\tau(C_j)| =\left\{\begin{array}{r@{\qquad}l} \geq 3 & {\rm
if~}j=1\\ 2 & {\rm otherwise}\end{array}\right.$
\smallskip

\begin{figure}[h]
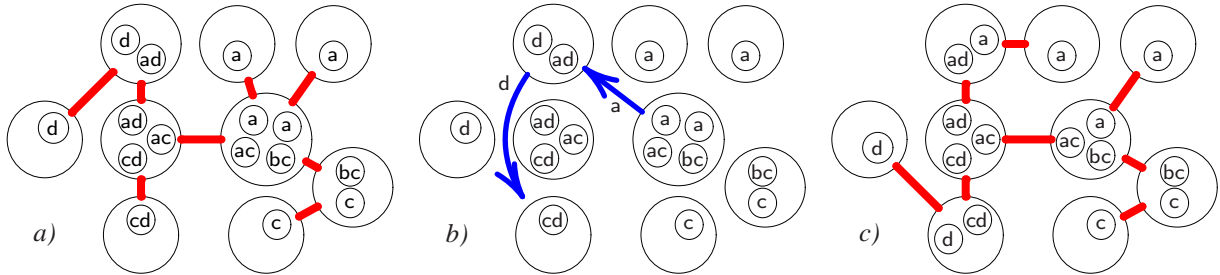

$$\xy/r3pc/:
(0,0)*++++[o][F]{\phantom{s}}="a";
"a"*{\xy/2pc/:
(0,0)*{};
(0.5,-0.5)*+[o][F]{\phantom{\bf _d}}="z";
"z"*{\sf _{d}};
\endxy};
(1,1)*+++++[o][F]{}="b";
"b"*{\xy/2pc/:
(0.4,0)*{};
(0,-0.5)*{};
(0,0.2)*+[o][F]{\phantom{\bf _4}}="z";
"z"*{\sf _{d}};
(-0.3,-0.2)*+[o][F]{\phantom{\bf _4}}="y";
"y"*{\sf _{ad}};
\endxy};
(1,0)*+++++[o][F]{}="c";
"c"*{\xy/2pc/:
(0,0.5)*{};
(-0.3,0.1)*+[o][F]{\phantom{\bf _4}}="z";
"z"*{\sf _{cd}};
(0,-0.35)*+[o][F]{\phantom{\bf _4}}="x";
"x"*{\sf _{ac}};
(0.3,0.1)*+[o][F]{\phantom{\bf _4}}="y";
"y"*{\sf _{ad}};
\endxy};
(1,-1)*++++[o][F]{\phantom{s}}="d";
"d"*{\xy/2pc/:
(0,0)*{};
(0.6,0)*+[o][F]{\phantom{\bf _4}}="y";
"y"*{\sf _{cd}};
\endxy};
(3,1)*++++[o][F]{\phantom{s}}="e'";
"e'"*{\xy/2pc/:
(0,0)*{};
(-0.6,0)*+[o][F]{\phantom{\bf _4}}="z";
"z"*{\sf _{a}};
\endxy};
(2,1)*++++[o][F]{\phantom{s}}="e";
"e"*{\xy/2pc/:
(0,0)*{};
(-0.6,0)*+[o][F]{\phantom{\bf _4}}="z";
"z"*{\sf _{a}};
\endxy};
(2.3,0)*+++++[o][F]{\phantom{s}}="f";
"f"*{\xy/2pc/:
(-0.5,0)*{};
(-0.2,0.35)*+[o][F]{\phantom{\bf _4}}="z";
"z"*{\sf _{ac}};
(0.2,-0.3)*+[o][F]{\phantom{\bf _4}}="x";
"x"*{\sf _{a}};
(0.3,0.22)*+[o][F]{\phantom{\bf _4}}="y";
"y"*{\sf _{a}};
(-0.3,-0.2)*+[o][F]{\phantom{\bf _4}}="w";
"w"*{\sf _{bc}};
\endxy};
(2.3,-1)*++++[o][F]{\phantom{s}}="g";
"g"*{\xy/2pc/:
(0,0)*{};
(0.45,-0.55)*+[o][F]{\phantom{\bf _4}}="y";
"y"*{\sf _{c}};
\endxy};
(3.2,-0.5)*+++++[o][F]{}="h";
"h"*{\xy/2pc/:
(0.4,0)*{};
(0,-0.5)*{};
(0.25,-0.2)*+[o][F]{\phantom{\bf _4}}="z";
"z"*{\sf _{bc}};
(-0.25,-0.2)*+[o][F]{\phantom{\bf _4}}="y";
"y"*{\sf _{c}};
\endxy};
{\ar@{-}@*{[red]}@*{[|<3pt>]} "a";"b"};
{\ar@{-}@*{[red]}@*{[|<3pt>]} "b";"c"};
{\ar@{-}@*{[red]}@*{[|<3pt>]} "c";"d"};
{\ar@{-}@*{[red]}@*{[|<3pt>]} "f";"e"};
{\ar@{-}@*{[red]}@*{[|<3pt>]} "f";"e'"};
{\ar@{-}@*{[red]}@*{[|<3pt>]} "c";"f"};
{\ar@{-}@*{[red]}@*{[|<3pt>]} "h";"g"};
{\ar@{-}@*{[red]}@*{[|<3pt>]} "f";"h"};
(0,-1)*{\mbox{\em a)}};
\endxy
\quad
\xy/r3pc/:
(0,0)*++++[o][F]{\phantom{s}}="a";
"a"*{\xy/2pc/:
(0,0)*{};
(0.5,-0.5)*+[o][F]{\phantom{\bf _d}}="z";
"z"*{\sf _{d}};
\endxy};
(1,1)*+++++[o][F]{}="b";
"b"*{\xy/2pc/:
(0.4,0)*{};
(0,-0.5)*{};
(0,0.2)*+[o][F]{\phantom{\bf _4}}="z";
"z"*{\sf _{d}};
(-0.3,-0.2)*+[o][F]{\phantom{\bf _4}}="y";
"y"*{\sf _{ad}};
\endxy};
(1,0)*+++++[o][F]{}="c";
"c"*{\xy/2pc/:
(0,0.5)*{};
(-0.3,0.1)*+[o][F]{\phantom{\bf _4}}="z";
"z"*{\sf _{cd}};
(0,-0.35)*+[o][F]{\phantom{\bf _4}}="x";
"x"*{\sf _{ac}};
(0.3,0.1)*+[o][F]{\phantom{\bf _4}}="y";
"y"*{\sf _{ad}};
\endxy};
(1,-1)*++++[o][F]{\phantom{s}}="d";
"d"*{\xy/2pc/:
(0,0)*{};
(0.6,0)*+[o][F]{\phantom{\bf _4}}="y";
"y"*{\sf _{cd}};
\endxy};
(3,1)*++++[o][F]{\phantom{s}}="e'";
"e'"*{\xy/2pc/:
(0,0)*{};
(-0.6,0)*+[o][F]{\phantom{\bf _4}}="z";
"z"*{\sf _{a}};
\endxy};
(2,1)*++++[o][F]{\phantom{s}}="e";
"e"*{\xy/2pc/:
(0,0)*{};
(-0.6,0)*+[o][F]{\phantom{\bf _4}}="z";
"z"*{\sf _{a}};
\endxy};
(2.3,0)*+++++[o][F]{\phantom{s}}="f";
"f"*{\xy/2pc/:
(-0.5,0)*{};
(-0.2,0.35)*+[o][F]{\phantom{\bf _4}}="z";
"z"*{\sf _{ac}};
(0.2,-0.3)*+[o][F]{\phantom{\bf _4}}="x";
"x"*{\sf _{a}};
(0.3,0.22)*+[o][F]{\phantom{\bf _4}}="y";
"y"*{\sf _{a}};
(-0.3,-0.2)*+[o][F]{\phantom{\bf _4}}="w";
"w"*{\sf _{bc}};
\endxy};
(2.3,-1)*++++[o][F]{\phantom{s}}="g";
"g"*{\xy/2pc/:
(0,0)*{};
(0.45,-0.55)*+[o][F]{\phantom{\bf _4}}="y";
"y"*{\sf _{c}};
\endxy};
(3.2,-0.5)*+++++[o][F]{}="h";
"h"*{\xy/2pc/:
(0.4,0)*{};
(0,-0.5)*{};
(0.25,-0.2)*+[o][F]{\phantom{\bf _4}}="z";
"z"*{\sf _{bc}};
(-0.25,-0.2)*+[o][F]{\phantom{\bf _4}}="y";
"y"*{\sf _{c}};
\endxy};
{\ar@/_1.5pc/@*{[blue]}@*{[|<2pt>]} "b";"d"};
{\ar@*{[blue]}@*{[|<2pt>]} "f";"b"};
"f"+(-0.64,0.34)*{_{\sf a}};
"b"+(-0.5,-0.4)*{_{\sf d}};
(0,-1)*{\mbox{\em b)}};
\endxy
\quad
\xy/r3pc/:
(0,0)*++++[o][F]{\phantom{s}}="a";
"a"*{\xy/2pc/:
(0,0)*{};
(-0.5,-0.5)*+[o][F]{\phantom{\bf _d}}="z";
"z"*{\sf _{d}};
\endxy};
(1,1)*+++++[o][F]{}="b";
"b"*{\xy/2pc/:
(0.4,0)*{};
(0,0.6)*{};
(0,-0.2)*+[o][F]{\phantom{\bf _4}}="z";
"z"*{\sf _{a}};
(-0.3,0.2)*+[o][F]{\phantom{\bf _4}}="y";
"y"*{\sf _{ad}};
\endxy};
(1,0)*+++++[o][F]{}="c";
"c"*{\xy/2pc/:
(0,0.5)*{};
(-0.3,0.1)*+[o][F]{\phantom{\bf _4}}="z";
"z"*{\sf _{cd}};
(0,-0.35)*+[o][F]{\phantom{\bf _4}}="x";
"x"*{\sf _{ac}};
(0.3,0.1)*+[o][F]{\phantom{\bf _4}}="y";
"y"*{\sf _{ad}};
\endxy};
(1,-1)*++++[o][F]{\phantom{s}}="d";
"d"*{\xy/2pc/:
(0,0)*{};
(0.4,-0.3)*{};
(0.6,0)*+[o][F]{\phantom{\bf _4}}="y";
"y"*{\sf _{cd}};
(0.3,0.45)*+[o][F]{\phantom{\bf _4}}="z";
"z"*{\sf _{d}};
\endxy};
(3,1)*++++[o][F]{\phantom{s}}="e'";
"e'"*{\xy/2pc/:
(0,0)*{};
(-0.6,0)*+[o][F]{\phantom{\bf _4}}="z";
"z"*{\sf _{a}};
\endxy};
(2,1)*++++[o][F]{\phantom{s}}="e";
"e"*{\xy/2pc/:
(0,0)*{};
(-0.6,0)*+[o][F]{\phantom{\bf _4}}="z";
"z"*{\sf _{a}};
\endxy};
(2.3,0)*+++++[o][F]{}="f";
"f"*{\xy/2pc/:
(0,-0.11)*{};
(0,0.75)*+[o][F]{\phantom{\bf _4}}="z";
"z"*{\sf _{ac}};
(0.3,0.27)*+[o][F]{\phantom{\bf _4}}="y";
"y"*{\sf _{a}};
(-0.2,0.27)*+[o][F]{\phantom{\bf _4}}="w";
"w"*{\sf _{bc}};
\endxy};
(2.3,-1)*++++[o][F]{\phantom{s}}="g";
"g"*{\xy/2pc/:
(0,0)*{};
(0.45,-0.55)*+[o][F]{\phantom{\bf _4}}="y";
"y"*{\sf _{c}};
\endxy};
(3.2,-0.5)*+++++[o][F]{}="h";
"h"*{\xy/2pc/:
(0.4,0)*{};
(0,-0.5)*{};
(0.25,-0.2)*+[o][F]{\phantom{\bf _4}}="z";
"z"*{\sf _{bc}};
(-0.25,-0.2)*+[o][F]{\phantom{\bf _4}}="y";
"y"*{\sf _{c}};
\endxy};
{\ar@{-}@*{[red]}@*{[|<3pt>]} "a";"d"};
{\ar@{-}@*{[red]}@*{[|<3pt>]} "b";"c"};
{\ar@{-}@*{[red]}@*{[|<3pt>]} "c";"d"};
{\ar@{-}@*{[red]}@*{[|<3pt>]} "b";"e"};
{\ar@{-}@*{[red]}@*{[|<3pt>]} "f";"e'"};
{\ar@{-}@*{[red]}@*{[|<3pt>]} "c";"f"};
{\ar@{-}@*{[red]}@*{[|<3pt>]} "h";"g"};
{\ar@{-}@*{[red]}@*{[|<3pt>]} "f";"h"};
(0,-1)*{\mbox{\em c)}};
\endxy
$$
\caption{{\em a)} token assignment $\tau$, {\em b)} augmenting path $(abc,adf,a)$,
$(adf,cdk,d)$ -- directed edges, {\em c)} $\tau$ after applying the path.
\label{fig:augmenting_path}}
\end{figure}

\noindent See Figure \ref{fig:augmenting_path} for an example of an augmenting
path of a token assignment $\tau$ and its application to $\tau$.

\begin{algorithm}[t]
\DontPrintSemicolon
\caption{\emph{Leafage($G$,$T$)}\label{alg:leafage}}
\KwIn{A chordal graph $G$, and a clique tree $T$ of $G$.}
\KwOut{A clique tree $T^*$ of $G$ with $|\Ll(T^*)|=\ell(G)$.}
\smallskip
Initialize $\tau \leftarrow \varepsilon_T$\tcc*[f]{initialize the token
assignment with the given clique tree.}\;
\While{there exists an augmenting path of $\tau$}{
	Let $(C_1, C_2, S_1), \ldots, (C_{k-1}, C_k, S_{k-1})$ be a shortest
	augmenting path of $\tau$\;
	\For{all $i$ from $1$ to $k-1$}{
		$\tau \leftarrow \tau \div (C_i, C_{i+1}, S_i)$\;
	}
} 
\Return{$T^*$ where $\varepsilon_{T^*} = \tau$.} \;
\end{algorithm}

It is easy to see that the application of an augmenting path decreases the
number of leaves in the resulting token assignment.  This, however, does not
guarantee that the resulting assignment corresponds to a clique tree of $G$.
Fortunately, it can be proved that a shortest augmenting path has this property,
and moreover, there always exists an augmenting path unless $\tau$ corresponds
to an optimal clique tree.  The details can be found in \cite{esa}.  We only
remark the following invariant which is maintaned throughout the algorithm.

\begin{lemma}{\rm\cite{esa}}\label{lem:inv}
In line 2 of Algorithm \ref{alg:leafage}, the variable $\tau$ is a realizable
token assignment.
\end{lemma}

After this introduction, we are ready to prove Theorem \ref{thm:3}.\bigskip

\begin{proofof}{Theorem \ref{thm:3}}
We prove the theorem by showing that each application of an augmenting path in
Algorithm \ref{alg:leafage} does not increase the number of leaves in the
subtrees of the corresponding tree model.

In other words, let $\tau$ be the token assignment considered at the start of
some iteration (Lines 2-5) of Algorithm~\ref{alg:leafage}, and let
$(C_1,C_2,S_1)$, $\ldots$, $(C_{k-1},C_k,S_{k-1})$ be the shortest augmenting
path of $\tau$ considered in this iteration (Line 3).  Let $\tau'$ denote the
value of $\tau$ after applying the token moves of this path (Lines 4-5).

By Lemma \ref{lem:inv}, both $\tau$ and $\tau'$ are realizable token assignments
of $G$. In other words, there exist clique trees $T$ and $T'$ of $G$ such that
$\tau=\varepsilon_T$ and $\tau'=\varepsilon_{T'}$.  Let
$\T_T=\big(T,\{T_u\}_{u\in V(G)}\big)$ and $\T_{T'}=\big(T',\{T'_u\}_{u\in
V(G)}\big)$ be the corresponding tree models of $G$. In other words, for each
$u\in V(G)$, we have $T_u=T\big[\{C\in V(T)~|~u\in C\}\big]$ and
$T'_u=T'\big[\{C\in V(T')~|~u\in C\}\big]$.  Moreover, just like in Lemma
\ref{lem:c1}, we define for each $u\in V(G)$ and each maximal clique $C$ of $G$,
the sets $\tau_u(C)=\{S~|~S\in\tau(C),u\in S\}$ and
$\tau'_u(C)=\{S~|~S\in\tau'(C),u\in S\}$. 

Now, to prove the theorem, it suffices to demonstrate that
$|\Ll(T_u)|\geq|\Ll(T'_u)|$ for every $u\in V(G)$.  Consider $u\in V(G)$ and
define two sequences of integers $a_1,\ldots,a_k$ and $b_1,\ldots,b_k$ where
$a_i=|\tau_u(C_i)|$ and $b_i=|\tau'_u(C_i)|$ for all $i\in\{1,\ldots,k\}$.  Note
that $\tau_u(C)=\tau'_u(C)$ for all $C\not\in\{C_1,\ldots,C_k\}$, and by Lemma
\ref{lem:c1}, we have $\Ll(T_u)=\big\{C~\big|~|\tau_u(C)|=1\big\}$ and
$\Ll(T'_u)=\big\{C~\big|~|\tau'_u(C)|=1\big\}$.  This implies the following.
\smallskip

$|\Ll(T_u)|-|\Ll(T'_u)|=\Big|\Big\{C_i~\Big|~|\tau_u(C_i)|=1\Big\}\Big|
-\Big|\Big\{C_i~\Big|~|\tau'_u(C_i)|=1\Big\}\Big| =
\Big|\{i~|~a_i=1\}\Big|-\Big|\{i~|~b_i=1\}\Big|$
\smallskip

\noindent In other words, this boils down to showing that $\{i~|~b_i=1\}$
does not have more elements than $\{i~|~a_i=1\}$.  

Recall that, by the definition of the augmenting path, $|\tau(C_k)|=1$ and
$|\tau(C_i)|=2$ for all $i\in\{2\ldots k-1\}$.  Notably, since the path is
shortest, $C_1,\ldots,C_k$ are distinct maximal cliques of $G$.  Thus, as
$\tau_u(C)\subseteq \tau(C)$ for all $C$, we conclude that $a_k\leq 1$ and
$a_i\leq 2$ for all $i\in\{2\ldots k-1\}$.  Further, note that $|\tau'(C_k)|=2$
while $|\tau'(C_i)|=|\tau(C_i)|=2$ for all $i\in\{2\ldots k-1\}$.  In other
words, we have $b_i\leq 2$ for all $i\in\{2\ldots k\}$. 

We shall use the following two claims to show that $\big|\{i~|~b_i=1\}\big|\leq
\big|\{i~|~a_i=1\}\big|$.

\begin{claim}\label{clm:thm3-1}
If $b_i=1$, then $a_i\geq 1$.
\end{claim}

\begin{proofclaim}
Consider $i\in\{1\ldots k\}$ such that $b_i=1$.  First, we show that $a_i\geq
1$. Suppose that $a_i=0$.  Since $b_i=1$, we have by Lemma \ref{lem:c1} that
$1=b_i=|\tau'_u(C_i)|={\rm deg}_{T'_u}(C_i)$.  In other words, $C_i$ is a leaf
of $T'_u$, and thus $T'_u$ contains at least 2 vertices. Recall that
$V(T_u)=V(T'_u)$, and note that $0=a_i=|\tau_u(C_i)|={\rm deg}_{T_u}(C_i)$ by
Lemma \ref{lem:c1}. This means that $C_i$ is a vertex of $T_u$ with no neighbour
in $T_u$. This is clearly impossible, since $T_u$ is connected and
$|V(T_u)|=|V(T'_u)|\geq 2$.  Thus we must conclude that $a_i\geq 1$.
\end{proofclaim}

\begin{claim}\label{clm:thm3-2}
If $b_i=1$ and $a_i\geq 2$, then there exists $j>i$ such that $a_j=1$, $b_j=2$, and
$a_{r}=b_r$ for all $r\in\{i+1,\ldots,j-1\}$.
\end{claim}

\begin{proofclaim}
To see this, first recall the construction of $\tau'$ from $\tau$ by
moving the tokens $S_1,\ldots,S_{k-1}$ as follows.  \medskip

\mbox{}\hfill$\tau'(C_i)=\left\{\begin{array}{l@{\qquad}l}
~~\,\tau(C_i)\setminus\{S_i\} & {\rm if~}i=1\\
\Big(\tau(C_i)\setminus\{S_i\}\Big)\cup\{S_{i-1}\} & {\rm if~}1<i<k\\
~~\,\tau(C_i)\qquad~\,\quad\cup\{S_{i-1}\} & {\rm if~}i=k
\end{array}\right.$\hfill\mbox{}
\medskip

Also recall that $a_i=\big|\tau_u(C_i)\big|=\big|\big\{S~|~S\in\tau(C_i),u\in
S\big\}\big|$ and $b_i=\big|\tau'_u(C_i)\big|=\big|\big\{S~|~S\in\tau'(C_i),u\in
S\big\}\big|$. From these two facts we conclude the following relationship
between the values of $a_i$ and $b_i$ ($1<i<k$).
\medskip

\noindent\mbox{}$(\star)$\hfill\quad
$b_1=\left\{\begin{array}{l@{\qquad}l}
a_1-1 & {\rm if~}u\in S_1\\
a_1 & {\rm if~}u\not\in S_1
\end{array}\right.$
\quad
$b_i=\left\{\begin{array}{l@{\qquad}l}
a_i & {\rm if~}u\in S_i\cup S_{i-1}\\
a_i-1 & {\rm if~}u\in S_i\setminus S_{i-1}\\
a_i+1 & {\rm if~}u\in S_{i-1}\setminus S_i\\
a_i & {\rm if~}u\not\in S_{i-1}\cup S_i
\end{array}\right.$
\quad
$b_k=\left\{\begin{array}{l@{\qquad}l}
a_k+1 & {\rm if~}u\in S_{k-1}\\
a_k & {\rm if~}u\not\in S_{k-1}
\end{array}\right.$\hfill\mbox{}\medskip

Now, for the proof of (\arabic{claim}), consider $i\in\{1\ldots k\}$ such that
$b_i=1$ and $a_i\geq 2$.  By ($\star$), we have $|b_i-a_i|\leq 1$ and thus
$a_i=2$.  Further, $i<k$ since $b_k\geq a_k$ by ($\star$), but
$b_i=1<2=a_i$.  Moreover, $u\in S_i$ since $i<k$ and $b_i=a_i-1$ by
($\star$).  We let $j$ be the largest in $\{i+1,\ldots,k+1\}$ such
that $a_r=b_r$ for each $r\in\{i+1,\ldots,j-1\}$.

First, we observe that $u\in S_r$ for each $r\in\{i,\ldots,j-2\}$.  Indeed, if
otherwise, we let $r$ be the smallest index in $\{i,\ldots,j-2\}$ with $u\not\in
S_r$.  As we just argued, we have $u\in S_i$, and so $r>i$.  Therefore, $u\in
S_{r-1}$ by the minimality of $r$. But then $b_r=a_r+1$ by ($\star$), since
$1\leq i<r<j-1\leq k$, a contradiction.

This also implies that $j\leq k$. Indeed, if $j=k+1$, then $i\leq j-2=k-1$ since
$i<k$. Thus $u\in S_{j-2}=S_{k-1}$ which yields $b_k=a_k+1$ by ($\star$).
However, $k\in\{i+1,\ldots,j-1\}$ and so $b_k=a_k$ by the choice of $j$.

We can now also conclude that $u\in S_{j-1}$. Indeed, if $i=j-1$, then we use
the fact that $u\in S_i$. Otherwise, $i\leq j-2$ in which case $u\in S_{j-2}$ as
argued above, and we conclude $u\in S_{j-1}$ by ($\star$), since $1\leq
i<j-1<k$.

Finally, we consider the value of $j$. First, suppose that $j=k$. Then
$b_k=a_k+1$, since $u\in S_{j-1}=S_{k-1}$. We recall that $a_k\leq 1$ and so
$b_k\in\{1,2\}$. If $b_k=1$, we have $a_k\geq 1$ by (\ref{clm:thm3-1}), but then
$a_k\geq b_k=a_k+1>a_k$, a contradiction. So, we must conclude $b_k=2$ and
$a_k=1$. Thus, as $j=k$, we have $b_j=2$, $a_j=1$, and $a_r=b_r$ for all
$r\in\{i+1,\ldots,j-1\}$ as required.  Thus we may assume that $j<k$. By the
maximality of $j$, we have $a_j\neq b_j$.  Also, $u\in S_{j-1}$ and $1\leq
i<j<k$. So by ($\star$) we conclude that $b_j=a_j+1$. We recall that
$b_j\leq 2$ as $j>1$. Thus $b_j\in\{1,2\}$ as $a_j\geq 0$. Again, if $b_j=1$, we
conclude $a_j\geq 1$ by (\ref{clm:thm3-1}) in which case $a_j\geq b_j>a_j$, a
contradiction. Thus $b_j=2$, $a_j=1$, and $a_r=b_r$ for all
$r\in\{i+1,\ldots,j-1\}$, as required.~\end{proofclaim}

We are now ready to conclude the proof.  Denote $A=\{i~|~a_i=1\}$ and
$B=\{i~|~b_i=1\}$. We show that $|B|\leq|A|$ which will imply the present
theorem as argued above the claim (\ref{clm:thm3-1}).

For each $i\in B$, if $a_i=1$, we define $\varphi(i)=i$; otherwise, we define
$\varphi(i)=j$ where $j$ is the index obtained by applying (\ref{clm:thm3-2})
for $i$; note that $a_j=1$ and $b_j=2$. It follows that $\varphi$ is a mapping
from $B$ to $A$. We show that $\varphi$ is, in fact, an injective mapping.
Suppose otherwise, and let $i,i^+$ be distinct elements of $B$ be such that
$\varphi(i)=\varphi(i^+)$. Recall that $b_i=b_{i^+}=1$ and note that $i\leq
\varphi(i)$ and $i^+\leq\varphi(i^+)$. If $i=\varphi(i)$, then $i^+\leq
\varphi(i^+)=\varphi(i)=i$ implying $i^+<\varphi(i^+)$ as $i$ and $i^+$ are
distinct. So $a_{i^+}\neq 1$ by the definition of $\varphi$, and hence
$b_{\varphi(i^+)}=2$ as $\varphi(i^+)$ was obtained by applying
(\ref{clm:thm3-2}) for $i^+$.  But then
$1=b_i=b_{\varphi(i)}=b_{\varphi(i^+)}=2$, a contradiction.  Thus we must
conclude that $i<\varphi(i)$ and, by symmetry, also $i^+<\varphi(i^+)$.  Now,
without loss of generality, assume $i<i^+$.  Since $i^+<\varphi(i^+)$, we must
have $a_{i^+}\neq 1$ by the definition of $\varphi$.  However, $b_{i^+}=1$ as
$i^+\in B$, and hence, $a_{i^+}\neq b_{i^+}$.  Recall that the choice of
$\varphi(i)$ using (\ref{clm:thm3-2}) for $i$ guarantees that $a_r=b_r$ for all
$r\in\{i+1,\ldots,\varphi(i)-1\}$. In particular,
$i<i^+<\varphi(i^+)=\varphi(i)$ and so $a_{i^+}=b_{i^+}$ which is a
contradition. This verifies that $\varphi$ is indeed an injective mapping from
$B$ to $A$, which yields $|B|\leq |A|$.

This concludes the proof of Theorem \ref{thm:3}.
\end{proofof}
\vspace{-2ex}

\section{Concluding Remarks}\label{sec:conclusion}

In this paper we have studied the vertex leafage of chordal graphs.
Specifically, a chordal graph $G=(V,E)$ has vertex leafage $k$ when it has a
tree model $\big(T,\{T_u\}_{u\in V}\big)$ such that each subtree $T_u$ has at
most $k$ leaves. We have shown that, for every fixed $k \ge 3$, it is
NP-complete to decide if a split graph $G$ has vertex leafage at most $k$ even
when $G$ is known to have vertex leafage at most $k+1$. Additionally, we have
demonstrated an $n^{O(\ell)}$ algorithm to compute the vertex leafage of a
chordal graph whose leafage is bounded by $\ell$. It remains open whether the
vertex leafage is FPT with respect to leafage (or any other graph parameter). 

Finally, we have shown that every chordal graph $G$ has a tree model which
simultaneously realizes $G$'s leafage and vertex leafage. In proving this final
result we have also shown that, for every path graph $G$, there exists a path
model with $\ell(G)$ leaves in the host tree and that such a path model can be
computed in $O(n^3)$ time.

\bibliographystyle{acm}
\bibliography{vertex-leafage}

\end{document}